\documentclass[twoside,11pt]{article}
\usepackage{amsmath,amssymb,amsfonts}
\usepackage{amsfonts}
\usepackage{graphicx}
\usepackage{graphics}
\usepackage{breqn}
\begin{document}
\setlength{\unitlength}{10mm}
%________________________Please insert your necessary newcommands____________________________________________
\newcommand{\f}{\frac}
\newtheorem{theorem}{Theorem}[section]
\newcommand{\sta}{\stackrel}
\def\be{\begin{equation}}
\def\ee{\end{equation}}
\def\bea{\begin{eqnarray}}
\def\eea{\end{eqnarray}}
\def\d{{\rm d}}
\def\e{\epsilon}
\def\m{\mu}
\def\n{\nu}
\def\mm{\mid}
\def\p{\partial}
\def\a{\alpha}
\def\t{\theta}
\def\g{\gamma}
%_________________________Pleae insert full name and address of authors______________________________________

\title{First class models from linear and nonlinear second class constraints }

\author{
Mehdi Dehghani \footnote{mdehghani@ph.iut.ac.ir, dehghani@sci.sku.ac.ir}
\\
{\scriptsize Department of Physics, Faculty of Science,
}\\{\scriptsize Shahrekord University, Shahrekord, P. O. Box 115, I. R. Iran}
\\
\\
Maryam Mardaani
\\
{\scriptsize Department of Physics, University of Kashan, Kashan, Iran.}
\\
\\
Majid Monemzadeh \footnote{monem@kashanu.ac.ir}
\\
{\scriptsize Department of Physics, University of Kashan, Kashan, Iran.}
\\
\\
Salman Abarghouei Nejad
\\
{\scriptsize Department of Physics, University of Kashan, Kashan, Iran.}}

%\\
%Full name of third author
%\\
%{\scriptsize Address of third Author} }

\date{}
\maketitle
\vspace{1cm}
%_____________________________________________________________________
%_________________________Please do not change this part any way____________________________________________
%\begin{picture}(5,2)(3,-9)
 %\put(0,1){{\scriptsize The Second Khansar Conference on Mathematical Sciences }}
 %\put(0,0.5){{\scriptsize $4^{th}$ Conference on Mathematical Analysis and its Applications }}
 %\put(0,0){{\scriptsize Faculty of Khansar, Khansar, Iran,  May 7-8, 2013}}
%\end{picture}
%_____________________________________________________________________
%_________________________Please type the abstract here_____________________________________________________
\vspace{-1.5cm}\begin{abstract}
\noindent  Two models with linear and nonlinear second class constraints are considered and gauged by embedding in an extended phase space. These models are studied by considering a free non-relativistic particle on the hyper plane and hyper sphere in the configuration space. The gauged theory of the first model is obtained by converting the very second class system to the first class one, directly.
In contrast, the first class system related to the free particle on the hyper sphere is derived with the help of the infinite BFT embedding procedure. We propose a practical formula, based on the simplified BFT method, which is practical in gauging linear and some nonlinear second class systems. As a result of gauging these two models, we show that in the conversion of second class constraints to the first class ones, the minimum number of phase space degrees of freedom for both systems is a pair of phase space coordinates.
This pair is made up of a coordinate and its conjugate momentum for the first model, but the corresponding Poisson structure of the embedded non-relativistic particle on hyper sphere is a non-trivial one. We derive infinite correction terms for the Hamiltonian of the nonlinear constraints and an interacting gauged Hamiltonian is constructed by summing over them. At the end, we find an open algebra for three first class objects of the embedded nonlinear system.

%\vspace{.5cm}\leftline{{AMS subject Classification 2010:
%Primary: 49M25; Secondary: 65D25, 65M70.}}

%\vspace{.5cm}
%\leftline{Keyword and phrases: {\small Pseudospectral },
%{\small Leg, }}

%\leftline{{\small  Differentiation }}
\end{abstract}
%_____________________________________________________________________

\newpage
\section{Introduction}
It seems that first and second class constrained systems are physically different. Although a first class constrained system is a gauge system, the second class one must be reduced to a physical system with physical degrees of freedoms. Unlike the second class systems, first class ones are more convenient to handle in the process of canonical quantization.

First class constraints, in the quantized version, acts on all states to select physical ones from many copies in the Hilbert or Fock space. On the other hand, second class constraints, when they are in the linear combinations of original phase space variables, can be removed in order to find physical phase space and construct Hilbert space in the quantization procedure.

 In order to quantize both systems, one must obtain the reduced phase space. The reduced phase space of a first class system is the selection of a plenty of similar points in the phase space that all of them satisfy first class identities. This selection is done with the help of other identities, considered by a gauge fixer. Gauge fixing conditions convert the set of first class constraints to second class one. In such a way, some degrees of the primary model are removed. In contrast, for second class systems, the reduced phase space is constructed by removing non-physical degrees of freedom corresponding to those constraints. For both first and second class systems the removing procedure is done by calculating Dirac brackets.

The gauge fixing process in theory of constrained systems is the key point to convert a second class model to a gauge model directly \cite{sarmadi} or conceptually \cite{BFT1,BFT2,BFT3}. One may imagine a set of second class constraints as a set of first class one and their gauge fixing conditions. The Dirac bracket removes additional degrees of freedom. Thus, in the reverse way, by adding some degrees of freedom to the second class functions, i.e. embedding the model in a extended phase space, we can convert them to a set of first class constraints. Because the decomposition of
a set of second class constraints into the first class one and gauge fixing conditions can be done in many ways, extracting gauge theory from a second class system has more than one solution. In this article, we see this point in the conversion of two sets of constraints. Although for the conversion of linear constraints we use a direct gauging process, for another one the famous method BFT \cite{BFT1,BFT2,BFT3}  rewritten for Abelian and
non-Abelian systems is used \cite{banerjee4}.

Somehow in a vise versa scheme, say converting first class constraints to second class ones, people  manipulate Hamiltonian constrained systems to enhance their symmetries and do quantization by modern and more mathematical technique named BRST method and de Rham cohomology \cite{taebook}. In this manner the phase space spreads by new variables with opposite Grassmannian number with respect to primary variables. See for a more related model to paper\cite{tae2}, which $SU(3)$ linear sigma model is considered.

In the present article we make a comparison between gauging a theory with linear constraints and gauging a theory with nonlinear ones. In this manner, in section (\ref{sec2}) we directly calculate the gauged model of a free non-relativistic particle on a hyper plane.
Although this model and its results are a part of our comparison, it teaches us the concept of gauging and embedding in a simple way.
Section (\ref{sec3}), which is the main part of current paper, is somehow a revision of gauging the Skyrme model or related models such as nonlinear sigma ($O(N)$ invariant) model, done with the help of the BFT embedding formalism. Most papers about these models are focused on the consistent canonical quantization and their quantum spectrum. This family of models were considered in several approaches including: the symplectic embedding
\cite{ananiasympl,soon3,barnetosym1,barwosym1}, the BFT formalism \cite{soon3,neto2, neto3,soon1,banerjee1,barcelos,barcelos2}, Stuckelberg field shifting \cite{neves,symplectic} or mixed approaches based on first principles of the making gauge systems \cite{soon3,symplectic,neto,soon2,deriglazov1}.

The problem stems from second class essence of the models, so people try to change it, quantize it and resile it. We want to know what kind of gauge theory can be constructed from such models, specially with the help of the BFT method, and what is their classical characteristics. To fulfil our goal, we clean the model from mess and limit it to the definite degrees of freedom to see the result, clearly. Extensions to more realistic models are not rigorous.
In part (\ref{sec3a}) we make a brief review on the BFT formalism. We reduce the formula of the BFT for our purposes to apply it on the model of free non-relativistic particle on a hyper sphere.
This model and the application of the BFT method on the model is introduced in (\ref{sec3b}). We find a general form of the embedded Hamiltonian of the free particle on the hyper sphere in (\ref{sec3b}) and make some comments on it in (\ref{sec3d}). We present the conclusions of our results in section (\ref{concl}).

\section{\label{sec2} Gauging by embedding a system with linear constraints}
Consider a free non-relativistic particle that its location is described by the coordinate $q_i$, which is displayed by a $D$-dimensional array $\vec{q}$. To have a second class constrained system we assume the particle is confined on a hyperplane. The mass of the particle play no rule in our analysis, so we scale the momenta by the mass of particle. The dynamics of such a system is described by the total Hamiltonian,
\begin{equation}\label{3a}
\begin{array}{ll}
 H_T = \frac{1}{2}\vec{p}.\vec{p}+\lambda\phi_1, &  \\
\phi_1 = \vec{a}.\vec{q} & \vec{a}.\vec{a}\neq0.
\end{array}
\end{equation}
Here $\vec{a}$ is a constant vector independent from phase space variables, which is normal of the hyperplane. with the help of this quantity we calculate the dynamics of $\phi_1$ and derive secondary constraint,
\begin{equation}\label{4}   \phi_2=\vec{a}.\vec{p}. \end{equation}
We see that the consistency of hyperplane in configuration space, as a primary constraint, leads us to another hyperplane in momenta sub-phase space with the same normal vector. The non-vanishing normal vector condition for the normal vector $\vec{a}$ makes primary and secondary constraints as the second class ones and truncates the consistency process of constraints.

Now, we transform our constraints to first class ones by extending the phase space. By a direct sum, we paste the auxiliary variables to the phase space and deduce the new phase space as follows,
\begin{equation}\label{5} \begin{array}{ll}  (q_i,p_i)\oplus(Q_j,P_j), & i=1,\dots D \hspace{.5cm} j=1,\dots d \\
\{q_i,p_{i'}\}=\delta_{ii'} & \{Q_j,P_{j'}\}=\delta_{jj'}. \end{array} \end{equation}
In the extended phase space we require that the constraints be corrected by new variables, linearly.
\begin{equation}\label{6}    \begin{array}{ll}
\widetilde{\phi_1}=\phi_1+\vec{b}.\vec{P}, & \widetilde{\phi_2}=\phi_2+\vec{c}.\vec{Q},  \end{array} \end{equation}
where $\vec{b}$ and $\vec{c}$ are two unknown vectors which are determined by two conditions. The first one is the first class condition as,
\begin{equation}\label{6a}    \{\widetilde{\phi_1},\widetilde{\phi_2}\}\approx0, \end{equation}
in which, the weak equality is the equality which is defined on the corrected constraints surface. The second condition is  deriving  $\widetilde{\phi_2}$ from the consistency of the $\widetilde{\phi_1}$. The later also required additional corrections to the Hamiltonian. It means that in the new phase space, the system is affected by a potential where for simplicity we assume it as a function $V=V(\vec{Q})$ in the new configuration space. So, we arrive to following equations.
\begin{equation}\label{7}    \begin{array}{l}  \vec{b}.\vec{c}=\vec{a}.\vec{a}, \\
\vec{c}.\vec{Q}+\vec{b}.\nabla_{\vec{Q}}V=0,   \end{array} \end{equation}
where, the $\nabla_{\vec{Q}}$ is the gradient operator with respect to $\vec{Q}$. It is worthwhile to note that there is a third condition which implies that after the investigating the consistency  of  $\widetilde{\phi_2}$ in the new model, no other constraint must be appeared.
\begin{equation}\label{8}    \{\phi_2,H_c\}+\widetilde{\lambda}\{\widetilde{\phi_1},\widetilde{\phi_2}\}\approx0. \end{equation}
In above equation both terms vanish identically. Hence, no new equation emerges. The first term comes from the characteristic of the primary model. The second one is due to the first classiness of the new model.

The set of partial differential equations (\ref{7}) have many solutions. A category of solutions can be derived by considering $\vec{b}$ and $\vec{c}$ as constant vectors. In this way, the primary free second class system (\ref{3a}) converts to the following interactive gauge system.
\begin{equation}\label{9}  \begin{array}{ll}
\widetilde{H_c}=\frac{1}{2}\vec{p}.\vec{p}-\frac{\vec{a}^2}{2\vec{b}^2}\vec{Q}.\vec{Q} &  \\   \widetilde{\phi_1}=\vec{a}.\vec{q}+\vec{b}.\vec{P} &
\widetilde{\phi_2}=\vec{a}.\vec{p}+\frac{1}{2}\frac{\vec{a}.\vec{a}}{\vec{b}.\vec{b}}\vec{b}.\vec{Q}.  \end{array}  \end{equation}
We see that the chain structure in the gauged model doesn't change, i.e. if we consider the $\widetilde{\phi_1}$
as the primary constraint, its consistency gives the $\widetilde{\phi_2}$ and the consistency of the later vanishes, strongly. Moreover, the algebra of the embedded constraints is an Abelian one.

As the normal vector of the constraint surfaces ($\vec{a}$) is characteristic of
the linear second class constrained model, the constant vector $\vec{b}$ which is incorporate with $\vec{a}$, is for the gauged system.
In addition, we find the surfaces that are described by the $(\widetilde{\phi_1}, \widetilde{\phi_2})$ are also another hyperplanes. But despite the presence of the constraint surfaces of the primary model in the coordinate and momentum sub-phase space, in the new model, the $(\widetilde{\phi_1}, \widetilde{\phi_2})$ are hyperplanes on the $\vec{q}\oplus\vec{P}$ and $\vec{p}\oplus\vec{Q}$ sub-phase spaces. Also, for primary model, normal vectors of the hyperplanes were the same, but for the new model they are different.

Moreover, There is a comment on the correction term which is added to the Hamiltonian. A physical interpretation for this term is that in spite of the free model, its gauged model is an interactive one. This describes an oscillator with incorrect sign in the potential. One may imagine that oscillator gives its energy from other part, according to the minus sign. This fact is understood better if one, by a canonical transformation, transforms the $\vec{Q}$ coordinates to momenta $\vec{P'}$.

Conclusively, we can select a minimal solution by considering only a pair of coordinate-momentum conjugate to gauge the free particle on a hyper surface as,
\begin{equation}\label{10}     \begin{array}{lll}
\widetilde{H_c}=\frac{1}{2}\vec{p}.\vec{p}-\frac{1}{2}\vec{a}.\vec{a}Q^2, & \widetilde{\phi_1}=\vec{a}.\vec{q}+P, &
\widetilde{\phi_2}=\vec{a}.\vec{p}+\vec{a}.\vec{a}Q.   \end{array} \end{equation}

For other second class constrained models, the extension of phase space and adding correction terms to the constraints and the Hamiltonian is not as simple as those models with linear constraints. There has been existed some conversion algorithms to gauge such systems. The BFT method is one of them which is used in the next section to convert the simplest model which contains nonlinear second class constraints.

\section{\label{sec3} Gauging by embedding a system with nonlinear constraints}

In this section we simplify the general form of the BFT algorithm to apply it on a model with nonlinear constraints. The present problem in its general form is related to the problem of the quantization of the free particle on sphere that is an introduction to the fundamental problem of the quantization in curved space-times. We focus only on the process of the conversion a classical second class system to a classical first class one. In other words, we work in the realm of pre-quantization of the systems with finite degrees of freedom and nonlinear second class constraints. Extension of the BFT formalism to infinite degrees (field) models is trivial but to the models with arbitrary nonlinearity of the constraints is not.

\subsection{\label{sec3a}BFT method}

The BFT algorithm, when it is applied on the systems with bosonic degrees of freedom, essentially comes from the conditions (\ref{5},\ref{6a}). Poisson structure of the adhered phase space to the primary phase space is not arbitrary but depends on the algebra of second class constraints. In this approach, to make a gauge theory, we have two sets of generators to generate correction terms for constraints and the Hamiltonian. They are denoted by the square
matrix $B$ and the vector $G$ for constraints and the Hamiltonian in the following relations, respectively. The correction terms are computed by,
\begin{eqnarray}  \nonumber  \phi_a^{(1)} &=& \chi_{ab}\eta_b, \\
\nonumber B_{ab}^{(1)} &=& \{\phi_a^{(0)},\phi_b^{(1)}\}-\{\phi_b^{(0)},\phi_a^{(1)}\}, \\
\nonumber B_{ab}^{(n)} &=& \sum^{n}_{m=0}\{\phi_a^{(n-m)},\phi_b^{(m)}\}+\sum^{n-2}_{m=0}\{\phi_a^{(n-m)},\phi_b^{(m+2)}\}_{(\eta)}, \\
\label{11} \phi_a^{(n+1)} &=& -\frac{1}{n+2}\eta_d\omega^{-1}_{dc}\chi^{-1}_{cb}B_{ba}^{(n)}, \hspace{5mm} n\ge1, \end{eqnarray}
for embedding the constraints and by
\begin{eqnarray}  \nonumber G^{(0)}_a &=& \{\phi_a^{(0)},H^{(0)}\}, \\
\nonumber G^{(1)}_a &=& \{\phi_a^{(1)},H^{(0)}\}+\{\phi_a^{(0)},H^{(1)}\}+\{\phi_a^{(2)},H^{(1)}\}_{\eta},\\
\nonumber G^{(n)}_a &=& \sum_{m=0}^{n}\{\phi_a^{(n-m)},H^{(m)}\} \\
\nonumber &+&\sum_{m=0}^{n-2}\{\phi_a^{(n-m)},H^{(m+2)}\}_{\eta}+\{\phi_a^{(n+1)},H^{(1)}\}_{\eta}, \\
\label{12} H^{(n+1)} &=&  -\frac{1}{n+1}\eta_a\omega^{-1}_{ab}\chi_{bc}G_{c}^{(n)}, \hspace{5mm} n\ge0 \end{eqnarray}
to gauge the Hamiltonian. The embedded Hamiltonian ($\widetilde{H}$) and the first class constraints ($\widetilde{\phi}$) are derived by applying the summation on corresponding correction terms. In above equations, the upper indices in the parentheses indicates the order of correction. Also, the $\eta_a$ is a vector which represents the new phase space variables, so the suffix $\eta$ under the brackets means the Poisson bracket in the adhered sub-phase space.
The quantities $\phi_a^{(0)}$ and $H^{(0)}$ are nothing more than uncorrected constraints and canonical Hamiltonian of the uncorrected model.
The Roman indexes $a,b,\cdots$ take their values from $\{1,2,\cdots, \sharp \textrm{of the second class constraints}\}$ and everywhere in this paper the summation convention is considered for repeated indices. The square matrices $\chi$ and $\omega$ determine how the elements of vector $\eta$ appear
in the correction terms. They satisfy the master equation of the BFT,
\begin{equation}\label{13}
 \Delta+\chi^{T}\omega\chi=0.
  \end{equation}

By determining $\omega$, also the Poisson structure of the new sub-phase space will be obtained, because we consider that there is no interaction between two parts of the phase space, i.e. it is off-shell.

As it has been stated before \cite{monemzade1,monemzade2}, there are more than one solution for equation(\ref{13}). Thus, for a given second class system, there are so many corresponding first class systems which divert to it after gauge fixing. In some cases, the elements of the matrix of Poisson brackets of the second
class constraints, say $\Delta_{ab}$, on the constraint surface are constant. Therefore, for such cases there is a simple solution for (\ref{13}) as one could assume the unknown matrices have elements independent from the phase space variables, even for the case $\chi=1$ and $\omega=-\Delta$. Such a solution reduces the recursion
relations (\ref{11},\ref{12}) to a simple form. In this regime the generating functions $B^{(n)}$ is vanished as same as Poisson brackets on new sub-phase space. Conclusively, the constraints are corrected by only one term, say $\widetilde{\phi_a}=\phi^{(0)}_a+\eta_a$. The recursion relation for the $n$th order of the Hamiltonian reduces to,
\begin{eqnarray}
\nonumber G^{(n)}_a &=& \{\phi^{(0)}_a,H^{(n)}\}, \\
H^{(n+1)} &=& \frac{1}{n+1}\eta_a\Delta^{-1}_{ab}G^{(n)}_b, \hspace{5mm} n\ge0.
\end{eqnarray}
The fractional factor can be absorbed in generating vector and the matrix $\Delta^{-1}$ rearranges the elements of $\vec{\eta}$, i.e. we can order the
recursion relations as,
\begin{eqnarray}
 \nonumber {G'}^{(n)}_a &=& \frac{1}{n+1}\{\phi^{(0)}_a,H^{(n)}\}, \\
\nonumber\eta'_b&=&\eta_a\Delta^{-1}_{ab}, \\
\label{15} H^{(n+1)} &=& \eta'_b {G'}^{(n)}_b, \hspace{5mm} n\ge0.
\end{eqnarray}

Although for the problem with which we have encountered, the $\Delta$ matrix does not have constant elements, we use the above simplified equations in an appropriate manner for our goal.

\subsection{\label{sec3b}BFT embedding a non-relativistic particle on hyper sphere}

The full dynamics of the free non-relativistic particle confined on a D-dimensional sphere is given by
\begin{equation}\label{1}
H_T=\frac{1}{2}\vec{p}.\vec{p}+\lambda (\vec{q}.\vec{q}-1),
\end{equation}
where we assume the confinement condition as a primary constraint. The $\lambda$ is Lagrange multiplier, adds the primary constraints $\phi_1$ to the canonical Hamiltonian. This is the simplest model with nonlinear constraint in the configuration space which produces its second class partner in the phase space as $\phi_2=\vec{q}.\vec{p}$ . Due to the nonlinear nature of the constraints, the extension of phase space in order to convert the constraints as gauge symmetries of a new model is not trivial. But in the BFT formalism we have sufficient equations to add a linear term to the constraints in order to make them first class. This assumption determines the symplectic structure of the extended phase  space. So, our consideration is that deformation of constraints by new phase space variables $(\eta_1,\eta_2)$ as $\widetilde{\phi_1}=\phi_1+\eta_1$ and $\widetilde{\phi_2}=\phi_2+\eta_2$ make them first class constraints.
Before starting the BFT embedding, the matrix elements of Poisson bracket of the constraints off the constraint surfaces is $\Delta_{ab}=2\vec{q}^2\e_{ab}$. During the  BFT process, the constraint surface changes, so we can't compute the $\Delta$ matrix on the constraint surface, unless up to the end of our calculations when it is corrected. This subtle point make the use of the (\ref{15}) problematic. We eliminate this problem by choosing a suitable ansatz, afterwards. In this way, according to the corrected constraints, we obtain the following nontrivial and nonconstant symplectic structure for the new part of the phase space,
\begin{equation}\label{1aa}
 \{\eta_a,\eta_b\}=-2(1-\eta_1)\e_{ab}.
 \end{equation}
The two dimensional antisymmetric tensor $\e_{ab}$ is characterized by $\e_{12}=1$.

We define the following objects to employ the relation (\ref{15}).
\begin{equation}\label{17}
\begin{array}{lll}
\overline{\phi}_0=H_c & \overline{\phi}_1=\phi_1+1 & \overline{\phi}_2=\phi_2.
\end{array}
 \end{equation}
One can shows, these quantities form a closed algebra with following structure constants.
\begin{eqnarray}
 \label{17a}\{\overline{\phi}_{\a},\overline{\phi}_{\beta}\} &=& f_{\a\beta\g}\overline{\phi}_{\g}, \\
\nonumber f_{\a\beta\g} &=&2 (\e_{\a\beta 0}\delta_{\g 1}+\e_{\a\beta 1}\delta_{\g 0}-\e_{\a\beta 2}\delta_{\g 2}),
 \end{eqnarray}
where Greek indices take their values from the set $\{0,1,2\}$. The $\delta_{\a\beta}$ and the  $\e_{\a\beta\g}$ are the conventional, 3-dimensional Kronecker discrete delta function and the full antisymmetric tensor with  $\e_{012}=1$, respectively. In this version, the three functions are first class in terminology of constrained systems. But they are not a full chain of a Hamiltonian and some constraints, specially $\overline{\phi}_1$ is not a constraint. In other words, the BFT process intends to induce a chain structure of Hamiltonian and constraints on these objects
by adding corrections to them.

By running the machinery of the simplified BFT (\ref{15}) for systems with constant $\Delta$ matrix (not only weakly but also off the constraint surface), one can deduce a general formula for $n$th order correction term of the embedded Hamiltonian, inductively.
\begin{equation}\label{18}
 H^{(n)}=\frac{1}{n!}\overline{\phi}_{\g_n}\prod^{n}_{m=1}\eta'_{a_m}f_{a_m\g_{m-1}\g_m}, \hspace{3mm}n\ge1,
  \end{equation}
where summation convention as before is considered for indices for their domain which is the set $\{1,2\}$ for Roman indices and the set $\{0,1,2\}$ for Greek indices. Also, the $\g_0$ takes only the value 0. Besides the current problem, we give an alternative approach to solve the linear problem in appendix (\ref{app A}) by the manipulation of the above instruction in the presence of central charges. This formula and its twin, that is given in (\ref{app A}), is one of the main results of this paper for gauging linear and nonlinear second class systems. Those are applicable and practical whenever someone tries to perform linear operations on second class functions to creates an algebra with the constant structure functions and central charges.

As we see in (\ref{18}), the $H^{(n)}$ can be expanded in terms of three elements of the closed algebra (\ref{17a}). In each level of iteration, due to the presence of $\Delta^{-1}$ that appears in the second equation of the (\ref{15}), also a factor $(\overline{\phi}_1)^{-1}$ is entered in $H^{(n)}$. Thus, the solution is not exactly linear in $\overline{\phi}_{\g}$ as (\ref{18}). Via afore thoughts, we guess the ansatz,
\begin{equation}\label{20}
   H^{(n)}=\frac{1}{(\overline{\phi}_1)^{n}}\overline{\phi}_{\m}F_{\m}(n;\vec{\eta}),
 \end{equation}
for the $n$th level of the correction to the Hamiltonian. After some calculation which appears in appendix (\ref{app B}) we arrive to the solutions,
\begin{eqnarray}   \nonumber F_{0}(n+1;\vec{\eta}) &=& (-\eta_1)^{n+1}, \\
\nonumber  F_{1}(1;\vec{\eta}) &=&0  \\   \nonumber F_{1}(n+1;\vec{\eta}) &=& \frac{1}{2}(\eta_2)^{2}(-\eta_1)^{n-1}, \\
\label{22}     F_{2}(n+1;\vec{\eta}) &=& \eta_2(-\eta_1)^n. \end{eqnarray}
Conclusively, the embedded Hamiltonian is a summation on all corrected terms plus the primary canonical Hamiltonian.
\begin{equation}\label{23}   \widetilde{H}=H_c+\sum_{n=1}^{\infty}H^{(n)}. \end{equation}
According to the equations (\ref{22}) and the ansatz (\ref{20}), the output is decomposed into three parts,
\begin{equation}\label{23a}
\widetilde{H}=\frac{1}{\overline{\phi}_1+\eta_1}(\overline{\phi}_1\overline{\phi}_0+\frac{1}{2}\eta^2_2+\eta_2\overline{\phi}_2).
\end{equation}
The convergence condition for the summations on the series which is appeared in the corrected
Hamiltonian, forces the value of the new phase space variable the limitation of $\eta_1<\frac{1}{2}$.
 Here is worth to noting that there is a Lagrangian approach which construct the whole gauged
 Hamiltonian without an infinite tower of correction terms \cite{derikuz,deribook}
\subsection{\label{sec3d} Verification of the results}
In the last stage of the gauging the model (\ref{1}), we verify whether our results are compatible or not. In this way, we check the Abelian or non-Abelian nature of the new first class constraints and their chain structure between themselves and the Hamiltonian. We purify the Hamiltonian in the form of kinetic and potential terms, by reduction of Hamiltonian on the new constraints surface.

In former subsection we see that due the Poisson structure of the new phase space  (\ref{1aa}), the corrected constraints are non-Abelian first class ones with the following algebra,
\begin{equation}\label{6aaa}
 \{\widetilde{\phi_1},\widetilde{\phi_2}\}=2\widetilde{\phi_1}.
\end{equation}

Although the first class constraints become Abelian in the finite order BFT, here we take out non-Abelian ones because of the non-constancy of the $\Delta$ matrix and infinite order of BFT project on this special problem.

In the next step, we consider that the $\widetilde{\phi_1}$  is a primary constraint for the system as like as its uncorrected partner which is described by $\widetilde{H}$. Afterwards, we simplify the Hamiltonian with the help of that to obtain
\begin{equation}\label{23aaa}
 \widetilde{H}'=\kappa H_c+\frac{1}{2}\varrho^2+\varrho\phi_2,
  \end{equation}
where we improve new variables by redefinitions,
\begin{equation}\label{30a}
\kappa=1-\eta_1, \hspace{5mm} \varrho=\eta_2.
\end{equation}
Then, the consistency of the $\widetilde{\phi_1}$ in the new system gives,
\begin{equation}\label{31a}
 \{\widetilde{\phi_1},\widetilde{H}'\}=2\varrho\widetilde{\phi_1}.
  \end{equation}
Which is vanished weakly and no another constraint will be emerged \footnote{Another possibility is that we encounter to a bifurcation in the process of consistency. But, by a direct calculation one can shows that the vanishing of the $\varrho$ eliminates the pair $(\kappa,\varrho)$ as a second class pair, which is not of our favorite.}. Consequently the second constraints situates at the primary level, inevitably. So, we project the Hamiltonian on the surface of both
constraints, named it $\widetilde{H}_{on}$, and then do the consistency. For the $\widetilde{\phi_2}$ it terminates to an identity due to the,
\begin{equation}\label{32a}   \{\widetilde{\phi_2},\widetilde{H}_{on}\}=0. \end{equation}
The curious reality that happen in this stage is that, if we project the $\widetilde{H}$ on the surface of both constraints (assume both of them as
primary constraints), then the chain structure remains according to the (\ref{6aaa}), (\ref{32a}) and
\begin{equation}\label{32aaa}  \{\widetilde{\phi_1},\widetilde{H}_{on}\}=2\kappa\widetilde{\phi_2}. \end{equation}
The three first class object make an open Lie algebra. Conclusively, after becoming first class, the chain structure of the constraints remains only on whole constraints surface. The completely on-shell Hamiltonian and its Poisson structure reduces to the absolute desired:
\begin{equation}\label{33a}   \begin{array}{ll}
\widetilde{H}_{on}=\frac{1}{2}\kappa\vec{p}^2-\frac{1}{2}\varrho^2,  & \kappa>\frac{1}{2} \\
\{\kappa,\varrho \}=2\kappa, & \{q_i,p_j\}=\delta_{ij} \\
\widetilde{\phi_1}=\vec{q}^2-\kappa & \widetilde{\phi_2}=\vec{q}.\vec{p}+\varrho.   \end{array} \end{equation}

At first, we see that the embedded hyper sphere in the configuration space doesn't transform to another hyper-sphere but it transforms to a hyper surface with sections, at $\kappa=constants$, in the form of hyper-sphere. It is a part of a spherical paraboloid. The minimum number of auxiliary variables in the primary model are imposed by construction of the BFT formalism, spontaneously.

The minus sign in front of the second term obliges us to interpret both $\kappa$ and $\varrho$ as coordinates, unless as same as the first example of this paper, we assume the primary system exchanges the energy with an external system in the form of an oscillator.  An extra evidence that guide us
to consider the $(\kappa,\varrho)$ as a coordinate pair is a non-canonical transformation
\begin{equation}\label{end1}     \begin{array}{ll} \kappa\rightarrow \ln\sqrt{\kappa}, & \varrho\rightarrow\varrho,    \end{array}  \end{equation}
which transforms this pair to usual canonical variables in the sense of usual symplectic structure on phase spaces \cite{sheikh}. So our consideration
is reasonable. If we get the $(\kappa,\varrho)$ as a pure coordinate pair, in quantization process, the new part of phase space is a noncommutative plane with position-dependent noncommutativity parameter. After the quantization is applied, it can be investigated in the context of Lie algebra noncommutativity \cite{Nair,Deriglazov3} or $\kappa$-Minkowski noncommutativity \cite{ruegg,camelia1,camelia2}.

\section{\label{concl}Conclusion}
In this paper we embed a general system with linear second class constraints directly with the help of the BFT embedding. We show that the embedded constraints also remain linear but the Hamiltonian becomes interactive. For a free particle on a hyper sphere, which can be considered as a model with nonlinear constraints, we do the embedding procedure by the finite order BFT formalism. But, because of the non-constancy of the $\Delta$ matrix, we
derive an infinite series for correction terms for the Hamiltonian.  In both systems, we find that correction terms for primary constraints are linear with respect to the new coordinates. It changes hyper sphere to the part of a spherical paraboloid. In conversion to first class systems we encounter non-unique solutions. This observation was seen by others in gauging another models in different approaches \cite{wotaz1,wotaz2,wotaz3}.  We find that minimal solutions add additional degrees of freedom in the number of the second class constraints, which brings to mind Wess-Zumino variables in gauging a system \cite{symplectic}. Our novelty of work is a prescription for computing the Hamiltonian correction terms, based  on the BFT formalism. Our general formula is derived for the fixed $\Delta$ matrix. We generalize it for nonconstant $\Delta$ matrix, i.e. off the  constraints surface. We find a nontrivial Poisson structure for the gauged nonlinear system, that in the quantization produces is a noncommutative plane. Ultimately, we show that after gauging the nonlinear system, the Hamiltonian and the constraints take place in a chain structure with non-Abelian open algebra.

\appendix

\section{\label{app A} }
For the linear problem we name the three objects of closed algebra as follow.
\begin{equation}\label{appa1}   \begin{array}{lll}
\overline{\phi}_0=H_c, & \overline{\phi}_1=\phi_1, & \phi_2=\overline{\phi}_2.    \end{array}  \end{equation}
All of the conventions for indices are as same as before. This algebra has an essential difference with respect to the algebra of the nonlinear problem.
In addition to structure constants it has a central charge in the form,
\begin{eqnarray}
\label{appa2}\{\overline{\phi}_{\a},\overline{\phi}_{\beta}\} &=& f_{\a\beta\g}\overline{\phi}_{\g}+c_{\a\beta}, \\
\nonumber f_{\a\beta\g} &=&\e_{\a\beta 2} \delta_{2\g}\\   \nonumber c_{\a\beta}&=&\vec{a}^2\e_{0\a\beta}. \end{eqnarray}
So, we are in the situation that we can generalize the (\ref{18}) in the presence of the central charges. In the same way which arrives us to the (\ref{18}), we can show that the compact term of the $H^{(n)}$ is decomposed into two parts,
\begin{eqnarray}\label{appa3}
 \nonumber H^{(n)}&=&\frac{1}{n!}(\overline{\phi}_{\g_n}\prod^{n}_{m=1}\eta'_{a_m}f_{a_m\g_{m-1}\g_m} \\
&+&c_{a_n\g_{n-1}}\eta'_{a_n}\prod^{n-1}_{m=1}\eta'_{a_m}f_{a_m\g_{m-1}\g_m}).
\end{eqnarray}
For $n=1$ the ambiguity in the second term of the parenthesis disappears, because according to the conventions and (\ref{appa2}), the $c_{a_0\g_0}=c_{a_0 0}$ vanishes.

In conclusion, we can establish the relations (\ref{appa2}, \ref{appa3}) to deduce the minimal solution (\ref{10}) after truncation at $n=2$. It means
that we have the finite order BFT which is due to the constant $\Delta$ matrix.
\section{\label{app B}}
We begin with the (\ref{15}) and the expression for the matrix elements of Poisson brackets of the second class constraints which can be written in the form $\Delta_{ab}=\frac{-1}{2{\overline{\phi}_1}}\e_{ab}$. In continue, we find the following recursion relation between consecutive corrections of the Hamiltonian.
\begin{equation}\label{app1}  H^{(n+1)}=\frac{-1}{2{\overline{\phi}_1}}\eta_a\e_{ab}\frac{1}{n+1}\{\overline{\phi}_b,H^{(n)}\}. \end{equation}
Afterward, we set the ansatz (\ref{20}) in the above equation to obtain
\begin{equation} \begin{array}{l}
\overline{\phi}_{\m} F_{\m}(n+1;\vec{\eta})=-\frac{1}{2}\eta_a\e_{ab}\frac{1}{n+1} (F_{\m}(n;\vec{\eta})f_{b\m\g} \overline{\phi}_{\g}   \\ \\
\hspace{28mm} -nF_{\m}(n;\vec{\eta})\overline{\phi}_{\m} \frac{1}{\overline{\phi}_1}f_{b1\g}\overline{\phi}_{\g}). \end{array} \end{equation}
The factor $\frac{1}{\overline{\phi}_1}$ in the second term is destroyer, but due to the special form of its coefficient $f_{b1\g}$ such a factor
is removed. It confirms the suggested ansatz and leads to the coupled recursion relations,
\begin{equation}\label{app3aaa}    \begin{array}{l}   F_0(n+1;\vec{\eta})=(-\eta_1)F_0(n;\vec{\eta}), \\
F_1(n+1;\vec{\eta}) =\frac{1}{n+1}(-(n-1)\eta_1F_1(n;\vec{\eta})+\eta_2F_2(n;\vec{\eta})), \\
F_2(n+1;\vec{\eta}) =\frac{1}{n+1}(\eta_2F_0(n;\vec{\eta})-n\eta_1F_2(n;\vec{\eta})),   \end{array}  \end{equation}
for three types of unknown functions $F_{\m}(n,\vec{\eta})$. The first and third equation of the above equations can be solved immediately up to
initial conditions $F_0(1;\vec{\eta})$ and $F_2(1;\vec{\eta})$. Then, with the help of these solutions we solve the second equation for $n\ge 1$. Eventually,
we read off the initial conditions and the $F_1(1;\vec{\eta})$ from $H^{(1)}$, directly. Putting them in the solutions,
\begin{equation}\label{app3aaa}   \begin{array}{l}
F_0(n+1;\vec{\eta})=(-\eta_1)^nF_0(1;\vec{\eta}),\\   \\   F_1(n+1;\vec{\eta}) =\frac{1}{2(n+1)}(-\eta_1)^{n-2}((n-1)\eta_2^2F_0(1;\vec{\eta})\\
\\ \hspace{18mm}-2\eta_1\eta_2F_2(0;\vec{\eta})),\\  \\
F_2(n+1;\vec{\eta}) =\frac{1}{n+1}(-\eta_1)^{n-1}(n\eta_2F_0(1;\vec{\eta})-\eta_1F_2(1;\vec{\eta})),     \end{array} \end{equation}
the answer (\ref{22}) will be obtained straightforwardly.

%\section*{References}


\begin{thebibliography}{0}
\bibitem{sarmadi} P. Mitra and R. Rajaraman, Ann. Phys. (N.Y.) {\bf 203}, 157, (1990).
  \bibitem{BFT1} I.A. Batalin and E.S. Fradkin,  Phys. Lett. {\bf B 180},  157,  (1986).
\bibitem{BFT2} I.A. Batalin and E.S. Fradkin, Nucl. Phys. {\bf B 279}, 514,  (1987).
\bibitem{BFT3} I.A. Batalin and I.V. Tyutin, Int. J. Mod. Phys. {\bf A 6}, 3255, (1991).
\bibitem{banerjee4} N. Banerjee, R. Banerjee and S. Ghosh, Ann. Phys. (N.Y.){\bf 241}, 237, (1994).
\bibitem{taebook} Soon-Tae Hong, in {\it BRST symmetry and de Rham cohomology.} (Springer, Heidelberg 2015).
\bibitem{tae2} S.-T. Hong,  S.H. Lee, Eur. Phys. J. {\bf C 25}, 131, (2002)
\bibitem{ananiasympl} J. Ananias Neto, C. Neves, W. Oliveira, Phys. Rev. {\bf D 63}, 085018, (2001).
\bibitem{soon3} S.-T. Hong, Y.-W. Kim, Y.-J. Park, K. Rothe,  J. Phys. {\bf A 36}, 1643, (2003).
\bibitem{barnetosym1} J. Barcelos-Neto, C. Wotzasek, Mod. Phys. Lett. {\bf A 7}, 1737, (1992).
\bibitem{barwosym1} J. Barcelos-Neto, C. Wotzasek, Int. J. Mod. Phys. {\bf A 7}, 4981 (1992).
\bibitem{neto2} W. Oliveira and J. Ananias Neto, Int. J. Mod. Phys. {\bf A 12}, 4895, (1997).
\bibitem{neto3} W. Oliveira and J. Ananias Neto, Nucl. Phys. {\bf B 533}, 611, (1998).
\bibitem{banerjee1} N. Banerjee, S. Ghosh, R. Banerjee, Nucl. Phys. {\bf B 417}, 257, (1994).
\bibitem{barcelos} J. Barcelos-Neto, Phys. Rev. {\bf D 55}, 2265, (1997).
\bibitem{barcelos2} J. Barcelos-Neto and W. Oliveira,  Phys. Rev. {\bf D 56}, 2257, (1997).
\bibitem{soon1} S.-T. Hong,  W.-T. Kim, Y.-J. Park, Phys. Rev. {\bf D 59}, 114026, (1999).
\bibitem{symplectic} J. Ananias Neto, Phys. Lett. {\bf B 571}, 105, (2003).
\bibitem{neves} C. Neves and C. Wotzasek, J. Phys. {\bf A 33}, 6447, (2000).
\bibitem{neto} E.M.C. Abreu, J. Ananias Neto, A.C.R. Mendes, C. Neves,  W. Oliveira, Ann. Phys {\bf 524}, 434, (2012).
\bibitem{soon2} S.-T. Hong, W.-T. Kim, Y.-J. Park, Phys. Rev. {\bf D 60}, 125005, (1999).
\bibitem{deriglazov1} A.A. Deriglazov, B.F. Rizzuti, Phys. Rev. {\bf D 83}, 125011, (2011).
\bibitem{monemzade1} M. Monemzadeh and A. Shirzad, Phys. Rev. {\bf D 72}, 045004, (2005).
\bibitem{monemzade2} M. Monemzadeh and A. Shirzad, Int. J. Mod. Phys. {\bf A 18}, 5613, (2003).
\bibitem{derikuz} A. A. Deriglazov and Z. Kuznetsova, Phys. Lett. {\bf B 646}, 47, (2007).
\bibitem{deribook} Alexei Deriglazov, in {\it Classical Mechanics Hamiltonian and Lagrangian formalism.} (Springer, 2010).
\bibitem{sheikh} M. Chaichian,  M.M. Sheikh-Jabbari, A. Tureanu,  Phys. Rev. Lett {\bf 86}, 2716, (2001).
\bibitem{Deriglazov3} A. A. Deriglazov, Phys. Lett. {\bf B 530}, 235, (2002).
\bibitem{Nair} V. P. Nair, Phys. Lett. {\bf B 505}, 249, (2001).
\bibitem{ruegg} J. Lukierski, H. Ruegg,  W.J. Zakrzewski, Ann. Phys {\bf 243}, 90, (1995).
\bibitem{camelia1} G. Amelino-Camelia, Phys. Lett. {\bf B 510}, 255, (2001).
\bibitem{camelia2} G. Amelino-Camelia, Int. J. Mod. Phys. Lett. {\bf D 11}, 35, (2002).
\bibitem{wotaz1} C. Wotzasek, Int. J. Mod. Phys. {\bf A 5}, 1123, (1990).
\bibitem{wotaz2} C. Wotzasek, J. Math. Phys. {\bf 32}, 540, (1991).
\bibitem{wotaz3} C. Wotzasek, Phys.Rev.Lett. {\bf 66}, 129,  (1991).

\end{thebibliography}
\end{document}